\documentstyle[multicol,aps,prb,epsf]{revtex}
\begin{document}

\draft

\title{
Possible high $T_c$ superconductivity mediated by 
antiferromagnetic spin fluctuations 
in systems with Fermi surface pockets
}

\author{
Kazuhiko Kuroki$^1$ and Ryotaro Arita$^2$
}

\address{$^1$ Department of Applied Physics and Chemistry, 
The University of Electro-Communications, 1-5-1 Chofugaoka, Chofu-shi,
Tokyo 182-8585, Japan\\
$^2$ Department of Physics, University of Tokyo, Hongo,Tokyo 113-0033, Japan}

\date{\today}

\maketitle

\begin{abstract}
We propose that 
if there are two small pocket-like Fermi surfaces, 
and the spin susceptibility is pronounced 
around a wave vector {\bf Q}
that bridges the two pockets, 
the spin-singlet superconductivity mediated by 
spin fluctuations may have a high transition temperature.
Using the fluctuation exchange approximation, 
this idea is confirmed for 
the Hubbard on a lattice with alternating hopping integrals, 
for which $T_c$ is estimated to be almost an order of magnitude
larger than those for systems with a large connected Fermi surface.
\end{abstract}

\medskip


\begin{multicols}{2}
\narrowtext
Fascination for electronic mechanisms of superconductivity 
has a long history. Given the large energy scale of the electrons itself, 
there has been an expectation from the early days that 
purely electronic mechanisms of superconductivity may lead to 
high transition temperature $(T_c)$.
The discovery of high $T_c$ superconductivity in the cuprates\cite{BM} 
has raised renewed interest in this possibility.
In fact, the repulsive Hubbard model, one of the simplest 
purely electronic models for the cuprates,
has been studied extensively to investigate whether the model 
can account for $T_c\sim 100$K.
In particular, the fluctuation exchange (FLEX) studies 
have shown that the Hubbard model near half-filling 
on a two dimensional (2D) square lattice exhibits 
spin-singlet $d$-wave superconductivity,
mediated by antiferromagnetic spin fluctuations, 
having $T_c\sim O(0.01t)$.\cite{Bickers}
Here $t$ is the hopping integral in the Hubbard model, 
which is estimated to be $\sim O(1eV)$ for the cuprates, 
so that $T_c\sim 100$ K, namely, a `high $T_c$'.
Studies along this line have been performed on various 
lattices near half-filling, where $T_c$ of the $d$-wave 
superconductivity has always 
turned out to be $\sim O(0.01t)$ on 2D\cite{Grabowski,Schmalian,KK1,KM} 
and quasi 1D lattices,\cite{KU,KK2,KK3} 
and even lower on 3D lattices.\cite{AKA}

On the other hand, the present authors have 
recently studied the Hubbard model
for various cases (i.e., combinations of 2D or 3D lattice structures and 
band filling) having {\em ferromagnetic} spin fluctuations to find 
that {\em spin-triplet} superconductivity mediated by 
ferromagnetic spin fluctuations, if any, has very low $T_c$ 
in general.\cite{AKA} 
In fact, the possibility of finite $T_c$ has been suggested only in 
systems having two {\em disconnected} pieces of Fermi surface  
lying point symmetrically about the $\Gamma$ point.\cite{KAr}
There, superconductivity is enhanced because the nodal lines 
of the gap function do not intersect the Fermi surface.

\begin{figure}
\begin{center}
\leavevmode\epsfysize=36mm \epsfbox{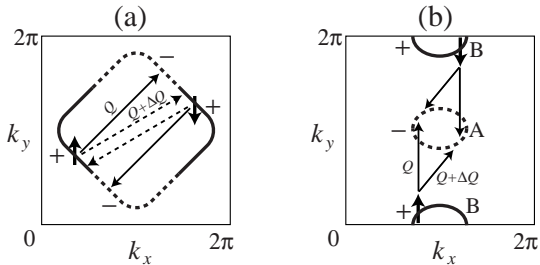}
\caption{
The $d$-wave gap function for a system with a large 
connected Fermi surface (a) or a gap function 
for systems with two pocket-like Fermi surfaces A and B (b).
The solid (dashed) curves represent positive (negative) values of the 
gap function on the Fermi surface. 
The solid (dashed) arrows represent spin-fluctuation-mediated 
pair scattering processes which give
positive (negative) contribution to $V_{\rm eff}$. 
{\bf Q} is the wave vector
at which the spin susceptibility peaks, while 
$\Delta${\bf Q} is the spread of the peak structure.
}
\label{fig1}
\end{center}
\end{figure}

Conversely, we may say that 
a reason why the singlet $d$-wave superconductivity
mediated by antiferromagnetic spin fluctuations 
has such `low $T_c$' $(\sim O(0.01t))$ compared to the 
original electron energy scale $t$ is  
because {\em the gap function has nodes on the Fermi surface}.
Let us first see why this is the case.
Superconductivity arises due to pair scattering processes
near the Fermi surface mediated by a pairing interaction.
Within the BCS theory, contributions to superconductivity 
from pair scattering processes
$[{\bf k, -k}]\in{\rm FS}\rightarrow [{\bf k', -k'}]\in{\rm FS}$ 
(FS stands for Fermi surface) are summed up in the form 
\[
V_{\rm eff}=
-\frac{\sum_{{\bf k,k'}\in {\rm FS}} 
V^{(2)}({\bf k-k'})\phi({\bf k})\phi({\bf k'})}
{\sum_{{\bf k}\in {\rm FS}}\phi^2({\bf k})},
\]
where $V^{(2)}({\bf k-k'})$ is the pairing interaction and 
$\phi({\bf k})$ is the superconducting gap function.
In order to have superconductivity with an order parameter $\phi$,
$V_{\rm eff}$ has to be positive and large. 
When the pairing is mediated by spin fluctuations that are pronounced 
around a certain wave vector {\bf Q}, 
the pairing interaction $V^{(2)}$ 
is positive and roughly proportional to the spin susceptibility,
so only the pair scattering processes 
$[{\bf k,-k}]\in{\rm FS}\rightarrow[{\bf k',-k'}]\in{\rm FS}$ 
that accompany a momentum transfer $\sim {\bf Q}$ and a sign change in $\phi$ 
give significant positive contributions to $V_{\rm eff}$.
This is the origin of the nodes in the $d$-wave gap function.
The nodes intersect the Fermi surface for a connected Fermi surface,
so that some pair scatterings give {\em negative} 
contributions to $V_{\rm eff}$ 
as far as the spin susceptibility has a finite spread $\Delta${\bf Q}
(see Fig.\ref{fig1}(a)).
In order to reduce negative contributions,
the spin susceptibility has to have a sharp structure around {\bf Q},
but in that case, pair scattering processes giving 
positive contributions will be strictly restricted to 
certain combinations of ${\bf k}$ and ${\bf k'}$,
so in any case, $V_{\rm eff}$ should be rather limited when the nodes of 
the gap function intersect the Fermi surface.

Due to the above considerations, we propose that 
if (i) there are two small pocket-like Fermi surfaces 
A and B and (ii) the spin susceptibility is pronounced 
around a wave vector {\bf Q} with a spread $\Delta${\bf Q},
where {\bf Q} {\em bridges} the Fermi surfaces A and B,
and $\Delta${\bf Q} is about the size of the pockets (see Fig.\ref{fig1}(b)), 
the resulting {\em spin-singlet} superconductivity 
may have a $T_c$ much larger than $0.01t$. Namely, 
with different signs of the gap function assigned 
\noindent
on A and B
(with no nodes on the Fermi surface), {\em interpocket} 
pair scattering processes 
$\forall[{\bf k,-k}]\in {\rm A}\rightarrow\forall[{\bf k',-k'}]\in {\rm B}$ 
(and vice versa) all 
have significant positive contribution to $V_{\rm eff}$,
resulting in a high $T_c$. 
This idea can be applied either to single band systems 
having disconnected pieces of Fermi surface or to 
systems with multiple Fermi surfaces corresponding to different bands.

\begin{figure}
\begin{center}
\leavevmode\epsfysize=50mm \epsfbox{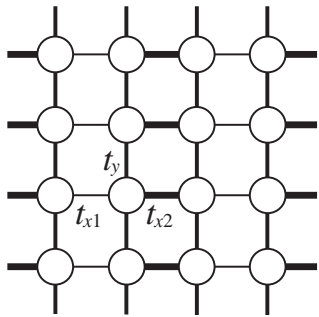}
\caption{The lattice considered in the present study.
}
\label{fig2}
\end{center}
\end{figure}

\begin{figure}
\begin{center}
\leavevmode\epsfysize=40mm \epsfbox{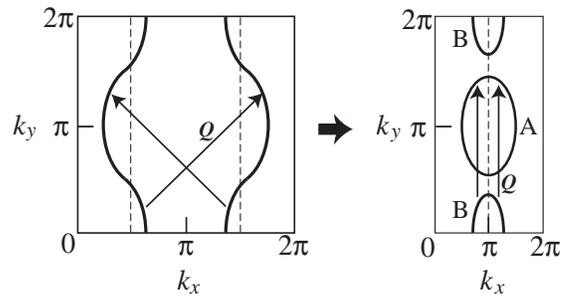}
\caption{The Brillouin zone and the 
Fermi surface(s) for $t_{x1}=t_{x2}$ (right panel)
and for $t_{x1}\neq t_{x2}$ (left panel).}
\label{fig3}
\end{center}
\end{figure}

In order to provide an example in which such a scenario is realized, 
we start with a 2D 
rectangular lattice with different hopping integrals between the 
$x$ and $y$ directions ($t_{x2}=t_{x1}\equiv t_x$ in Fig.\ref{fig2}). 
For $t_x/t_y>1$, the Fermi surface becomes open in the $y$ direction 
(Fig.\ref{fig3},left panel). When the band filling is close to 
half-filling, i.e. $n\sim 1$, the Fermi surface intersects 
the lines $k_x=\pi,3\pi/2$ for values of $t_x/t_y$ not too 
large.\cite{comm1D}
The nesting vector {\bf Q} in this case bridges the parts of the 
Fermi surface that lie inside and outside of the region
$\pi/2<k_x<3\pi/2$. If we now introduce an alternation 
in the hopping integrals in the $x$ direction 
($t_{x1}\neq t_{x2}$ in Fig.\ref{fig2}), the band  
will be folded at $k_x=\pi,3\pi/2$, resulting in two pocket-like
Fermi surfaces, each in different bands (Fig.\ref{fig3},right panel). 
The important point 
here is that the nesting vector {\bf Q} should not change, at least 
when $t_{x1}/t_{x2}$ is not too far from unity, so that 
{\bf Q} {\em bridges the two Fermi surfaces}. 

We have performed FLEX calculation for the Hubbard model,
${\cal H}=-\sum_{\langle i,j \rangle \sigma=\uparrow,\downarrow} 
t_{ij}(c^{\dagger}_{i\sigma}c_{j\sigma}+c^{\dagger}_{j\sigma}c_{i\sigma})
+U\sum_i n_{i \uparrow}n_{i \downarrow}$,
on the above lattice. 
In the two band version of FLEX,\cite{KU} the Green's function $G$, 
the susceptibility $\chi$, the self-energy $\Sigma$, and 
the superconducting gap function $\phi$ all become $2\times 2$ matrices,
e.g., $G_{lm}({\bf k}, i\varepsilon_n)$, where $l,m$ specify
the two sites in a unit cell.
The orbital-indexed matrices for Green's function and the gap functions
can be converted into band(denoted as $\alpha$ and $\beta$)-indexed 
ones with a unitary transformation. 
As for the spin susceptibility, we diagonalize the 
$2\times 2$ matrix $\chi_{\rm sp}$ and 
concentrate on the larger eigenvalue, denoted as $\chi$.  

The actual calculation proceeds as:
(i) Dyson's equation is solved to obtain the 
renormalized Green's function $G(k)$, 
where $k$ is a shorthand for the wave vector ${\bf k}$ and 
the Matsubara frequency, $i\epsilon_n$; 
(ii) The effective interaction 
$V^{(1)}(q)$ is given as
$V^{(1)}(q)=\frac{3}{2}V_{\rm sp}(q)+\frac{1}{2}V_{\rm ch}(q)
-U^2\chi_{\rm irr}(q)$.
Here, 
the effective interactions due to 
spin fluctuations (sp) and those due to charge 
fluctuations (ch) have the forms
$V_{\rm sp}=U^2\chi_{\rm sp}$ and 
$V_{\rm ch}=U^2\chi_{\rm ch}$, respectively, 
where the spin and the charge susceptibilities are 
$\chi_{\rm sp(ch)}(q)=\chi_{\rm irr}(q)[1-(+)U\chi_{\rm irr}(q)]^{-1}$
in terms of the irreducible susceptibility matrix 
$\chi_{\rm irr}(q)= -\frac{1}{N}\sum_k G(k+q)G(k)$
($N$:number of $k$-point meshes);
(iii) $V^{(1)}$ then brings about the self-energy, 
$\Sigma(k)=\frac{1}{N}\sum_{q} G(k-q)V^{(1)}(q)$,
which is fed back to Dyson's equation, 
and the self-consistent iterations are 
repeated until convergence is attained.
We take $64\times 64$ $k$-point meshes and 
up to 4096 Matsubara frequencies. 

We determine $T_c$ as the temperature at which the eigenvalue $\lambda$ of 
the Eliashberg equation, 
\begin{eqnarray*}
\lambda\phi_{l m}(k)\nonumber &=& -\frac{T}{N}\sum_{k'}\sum_{l',m'}\\
&&\times V_{l m}^{(2)}(k-k')G_{ll'}(k')G_{mm'}(-k')\phi_{l'm'}(k'),
\end{eqnarray*}
reaches unity.
Here the pairing interaction $V^{(2)}$ for singlet pairing 
is given by $V^{(2)}(q)=\frac{3}{2}V_{\rm sp}(q)-\frac{1}{2}V_{\rm ch}(q)$.

We take $t_y$ as the unit of the energy $(t_y=1)$.
We present the results in the multiband scheme even 
for the single band case of $t_{x1}=t_{x2}$. 
Throughout the study, the band filling (number of electrons / number of sites)
and the on-site repulsion are fixed at $n=0.95$ and $U=7$, respectively. 

\begin{figure}
\begin{center}
\leavevmode\epsfysize=55mm \epsfbox{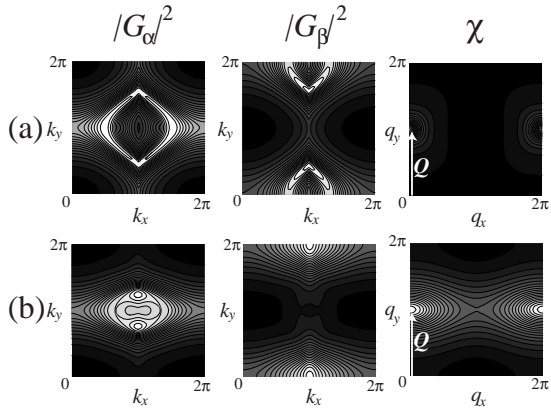}
\caption{Contour plots of $|G_\alpha({\bf k})|^2$, 
$|G_{\beta}({\bf k})|^2$ and $\chi({\bf q})$ 
for $t_{x1}=$1.3(a) and 0.8(b) with $t_{x2}=1.3$ and $T=0.14$. 
The lighter the color, the larger the value.
Note that $|G_\alpha|^2$ in (b) is a contour plot of 
Fig.\protect\ref{fig7}(c).}
\label{fig4}
\end{center}
\end{figure}

Let us now proceed to the calculation results. 
We first vary $t_{x1}$ fixing $t_{x2}=1.3$ (solid arrow in the inset of
Fig.\ref{fig6}(a)).
In Fig.\ref{fig4}, we plot $|G_\alpha|^2$, $|G_\beta|^2$ and $\chi$
for the lowest Matsubara frequency for $t_{x1}=1.3$ (a) and $0.8$ (b).
The ridges of $|G|^2$ represent the Fermi surfaces.
For $t_{x1}=1.3(=t_{x2})$, this is a single band system, so the 
Fermi surfaces in the $\alpha$ and $\beta$ bands are connected at around
$k=(\pi,\pi/2)$ and $(\pi,3\pi/2)$ to result in 
a single, large (open) Fermi surface. The spin susceptibility 
peaks at the nesting vector of the Fermi surface, namely at ${\bf Q}=(0,\pi)$
in the folded Brillouin zone (or ${\bf Q}=(\pi,\pi)$ 
in the unfolded Brillouin zone).
The corresponding gap function for the lowest Matsubara frequency
shown in 

\begin{figure}
\begin{center}
\leavevmode\epsfysize=70mm \epsfbox{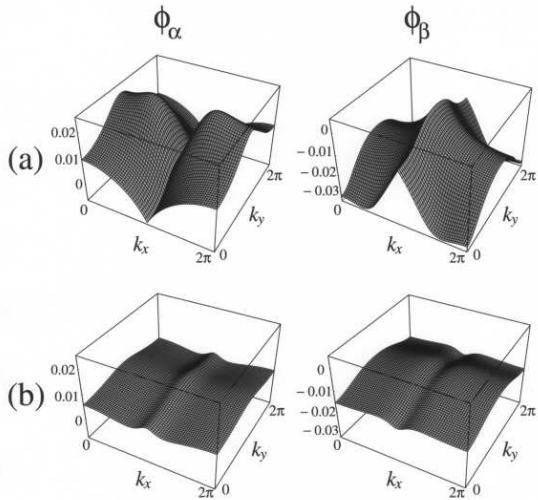}
\caption{Plots of $\phi_\alpha({\bf k})$ and $\phi_\beta({\bf k})$ 
for the same sets of parameter values as in Fig.\protect\ref{fig4}.}
\label{fig5}
\end{center}
\end{figure}

\begin{figure}
\begin{center}
\leavevmode\epsfysize=80mm \epsfbox{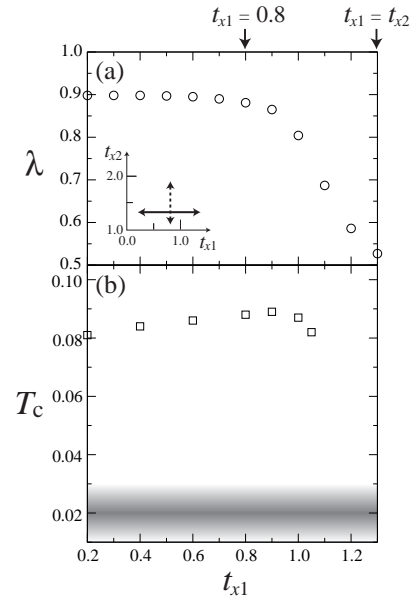}
\caption{$\lambda$(a) and $T_c$(b) plotted 
as functions of $t_{x1}$ with $T=0.14$ 
in (a) and $t_{x2}$ fixed at 1.3. $T_c$ is not obtained for $t_{x1}> 1.1$,
where the system comes very close to antiferromagnetic ordering 
($U\chi_{\rm irr}({\bf Q})\sim 1$) before $\lambda$ reaches unity
upon lowering the temperature. The hatched region in (b) represent 
the range of $T_c$ for systems having a large connected Fermi surface. 
The solid(dashed) arrow in the inset of (a) shows the range of the 
parameter values taken in Fig.\protect\ref{fig6}(Fig.\protect\ref{fig8}).
}
\label{fig6}
\end{center}
\end{figure}
\noindent
Fig.\ref{fig5}(a) has a node along $k_x\sim\pi$ in order to
satisfy the condition $\phi({\bf k})=-\phi({\bf k+Q})$ 
on the Fermi surface.

For $t_{x1}=0.8$ by contrast, there is a small pocket-like Fermi surface 
in each band, namely, around ${\bf k}=(\pi,\pi)$ in the $\alpha$ band 
and around $(\pi,0)$ in the $\beta$ band. 
The spin structure becomes broader having a spread $\Delta${\bf Q} of the
size of the pockets, and it still peaks around ${\bf Q}=(0,\pi)$,
so that ${\bf Q}$ {\em bridges} the two Fermi surfaces. 
As seen in Fig.\ref{fig5}(b), the corresponding gap functions 
have fixed signs in each band, and their variation is small.
Thus, interpocket pair scatterings 
$\forall[{\bf k,-k}]\in {\rm A}\rightarrow\forall[{\bf k',-k'}]\in {\rm B}$
(and vice versa) all have significant positive contributions to $V_{\rm eff}$.
 
Correspondingly, as shown in Fig.\ref{fig6}(a),
the maximum eigenvalue $\lambda$ for $T=0.14$ 
starts out around $\sim 0.5$ for the single-band case
of $t_{x1}=t_{x2}=1.3$, but 
it increases up to $\sim 0.9$ upon decreasing $t_{x1}$. 
$T_c$, plotted in Fig.\ref{fig6}(b), comes close to $\sim 0.1$
for $t_{x1}<1$, which is almost an order of magnitude larger 
compared to typical values 
for the cases with a large connected Fermi surface.

In order to show that the topology and the size of the Fermi surfaces are  
crucial, we now show results for various values of $t_{x2}$,
fixing $t_{x1}=0.8$ (dashed arrow in the inset of
Fig.\ref{fig6}(a)). When $t_{x2}\simeq t_{x1}$ or smaller,
the Fermi surfaces in both bands are open in the $x$ direction.
This 
\begin{figure}
\begin{center}
\leavevmode\epsfysize=90mm \epsfbox{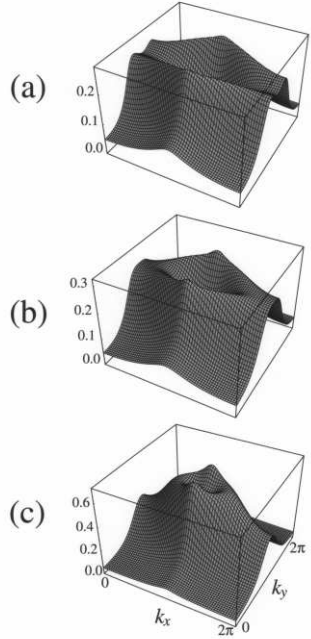}
\caption{Plots of $|G_{\alpha}({\bf k})|^2$ 
for $t_{x2}=$1.1(a), 1.2(b), and 1.3(c) 
with $t_{x1}=0.8$ and $T=0.14$.}
\label{fig7}
\end{center}
\end{figure}

\begin{figure}
\begin{center}
\leavevmode\epsfysize=40mm \epsfbox{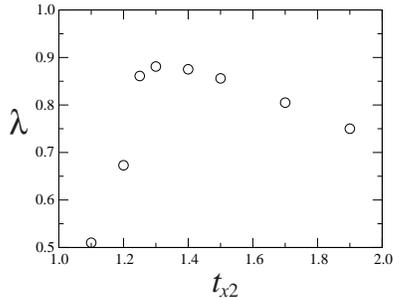}
\caption{$\lambda$ plotted 
as a function of $t_{x2}$ with $t_{x1}=0.8$ and $T=0.14$.}
\label{fig8}
\end{center}
\end{figure}
\noindent
can be understood by considering the single band situation
$t_{x2}=t_{x1}<t_y$, where the unfolded Fermi surface for $n\sim 1$ 
is open in the $x$ direction.
On the other hand, we have seen that 
there are two small pocket-like Fermi surfaces for $t_{x2}=1.3$,
so each Fermi surface has to change 
its topology between $t_{x2}\sim 0.8$ and $1.3$.
In fact, it can be seen from Fig.\ref{fig7} that the topology of the 
Fermi surface in the $\alpha$ band changes from a large Fermi surface
to a small pocket-like Fermi surface at around $t_{x2}\sim 1.2$.
$\lambda$ rapidly increases corresponding to this change of the 
topology of the Fermi surface, as can be seen from Fig.\ref{fig8}.
On the other hand, $\lambda$ decreases for too large $t_{x2}$ because
the $\beta$ band lies above the Fermi level in this case.\cite{commBW}

Finally, let us discuss the relation between the present study and 
previous ones.
The Hubbard model on coupled-ladder lattices\cite{KU,KK3} has been 
studied using FLEX approximation.  In those studies, 
$T_c\sim O(0.01t)$ has been obtained, which, 
in light of the present results, is because the 
parameter values adopted there do not result 
in small, pocket-like Fermi surfaces. 

In the large $U$ limit, the present Hubbard model should tend to the  
$t$-$J$ model with alternating $t$ and $J$.
The $t$-$J$ model with alternating $J$ (dimerized $t$-$J$ model) 
has been studied as a {\em spin gapped} system.\cite{Imada} 
There, however, the correlation between superconductivity 
and the topology/size of the Fermi surfaces was not discussed.
It is not clear at present whether the $t$-$J$ model with
alternation only in $J$ has properties similar to those presented here.
The Hubbard and $t$-$J$ models on {\em purely} 
1D ladder lattices have also been 
studied extensively as spin gapped systems.
In particular, large enhancement of the pairing correlation 
has been obtained when the Fermi level lies near the bottom of the upper band
\cite{Yamaji,Noack,KA}. These results seem to be related to the present 
study, although we cannot make a definite conclusion because the 
FLEX approximation cannot be applied to purely 1D systems, which are  
not Fermi liquid but Luttinger liquid.

At present, it is not clear whether the presence of 
a spin gap is a sufficient or a necessary condition for a
superconducting state without nodes of the gap on the Fermi surface.
Intuitively, however, a gapfull superconductivity and a finite 
spin gap do seem to be consistent with each other.
This point remains as an appealing future problem.

Discussions with Hideo Aoki and Takashi Kimura are gratefully acknowledged.  
This work is in part funded by the Grant-in-Aid for Scientific
Research from the Ministry of Education of Japan.

\end{multicols}

\begin{references}
\bibitem{comm1D} In the purely 1D limit, that is, 
when $t_x/t_y\rightarrow\infty$, the Fermi surface becomes a 
straight line parallel to the $y$ axis, so it does not
intersect the lines $k_x=\pi,3\pi/2$ for $n\neq 1$.
\bibitem{commBW} For $t_{x2}$ as large as $\sim2$, 
the band width is larger than the case 
for $t_{x1}=t_{x2}=t_y=1$, so that we are actually dealing with an effectively 
lower temperature. For $t_{x2}\sim 1.3$ and $t_{x1}\sim 0.8$, however,
the band width is similar to that for $t_{x1}=t_{x2}=t_y=1$, so our 
conclusion that $T_c$ is almost an order of magnitude larger in the former 
compared to the latter holds.
\end{references}
\end{document}